\documentclass[9pt,twocolumn,twoside]{opticajnl}
\journal{opticajournal} % use for journal or Optica Open submissions

\setboolean{shortarticle}{true}
\usepackage{color,soul, xcolor}

\usepackage{float}

\usepackage{lineno}
\title{Broadband Optoelectronic Mixer for Terahertz Frequency-Comb Measurements}

\author[1,2,*]{Jizhao Zang}
\author[3]{Jesse S. Morgan}
\author[3]{Andreas Beling}
\author[1,2]{Scott B. Papp}

\affil[1]{Time and Frequency Division, National Institute of Standards and Technology, Boulder, Colorado, USA}
\affil[2]{Department of Physics, University of Colorado, Boulder, Colorado, USA}
\affil[3]{Department of Electrical $\&$ Computer Engineering, University of Virginia, Charlottesville, Virginia, USA}

\affil[*]{jizhao.zang@nist.gov}

\begin{abstract}
We demonstrate ultra-broadband optoelectronic mixing of frequency combs that provides phase-coherent detection of a repetition frequency up to 500 GHz, using a high-speed modified uni-traveling carrier (MUTC) photodiode. Nonlinear photo-electron effects in the photodiode itself enable harmonic generation and down-mixing process of combs with widely different repetition frequency. Specifically, we generate two, 25 GHz frequency combs and use an optical filter to explore coherent down-mixing to baseband of comb spectral components across microwave, millimeter wave, and terahertz (THz) frequencies. The exceptional noise performance of the optoelectronic mixer enables the phase-coherent measurement of millimeter-wave and THz frequency combs with an Allan deviation of $ 10^{-13}/\tau$ for a measurement time of $\tau$. We further investigate the dependence of conversion loss on the reverse bias voltage and photocurrent. The experimental results indicate that we can minimize the conversion loss by operating the photodiode at an optimal voltage and maximum available photocurrent. Our work provides a solution for millimeter-wave and THz frequency comb measurements and facilitates fully stabilized frequency combs with microresonators. 
\end{abstract}

\setboolean{displaycopyright}{false} % Do not include copyright or licensing information in submission.

\begin{document}
\maketitle
%\scott{The premise of this paper is:  photomixing, aided by nonlinearity, of two frequency combs on a detector in order to measure a high repetition frequency}

Photonic integration of optical-frequency synthesizers \cite{Daryl2018} and optical-clock dividers \cite{Tara2019}, which generate laser sources with an optical frequency programmed from a microwave frequency reference and generate microwave (MW) signals from an optical clock, respectively, have the potential to revolutionize a wide range of applications from signal generation to information processing. The synthesis of optical frequencies relies on an optical frequency comb, which consists of equally spaced modes at frequencies $\nu_n=f_{\text{ceo}}+n\times f_{\text{rep}}$, where $f_{\text{ceo}}$ is the carrier-envelope offset frequency, $n$ is the mode number, and $f_{\text{rep}}$ is the repetition rate. Recently, this field has been further advanced by the introduction and use of microresonator-based optical frequency combs (microcombs) \cite{TobiasK2018}. Featuring compact size, flexible dispersion engineering, and low power consumption, microresonator combs offer the opportunity to realize a single-chip solution for optical-frequency metrology. 

Optical frequency synthesis and optical-clock division require fully stabilized octave-spanning microcombs\cite{Briles2018, Tara2019}. But to date, most examples of octave span microcombs operate with high repetition frequency in the THz range of about 100 GHz to 1000 GHz, which is challenging to phase-coherently measure with respect to a microwave frequency reference. High-repetition-frequency microcombs are preferred because of lower threshold power and larger mode spacing that facilitates extension of the comb span. To measure such large $f_{\rm{rep}}$, experiments have used a microwave rate ( $f_{\text{mw}}$) auxiliary frequency comb \cite{Briles2018} or electro-optic modulation \cite{Jordan2018} for down conversion. By heterodyne detection of the two combs, we can down-convert $f_{\text{rep}}$ to an intermediate frequency (IF) signal $f_{\text{IF}}$ since $f_{\text{rep}}=Nf_{\text{mw}}+f_{\text{IF}}$, where $N$ is an integer. However, these types of schemes are complicated by the relative optical phase of the two combs. 

We anticipate that by photodetecting frequency combs of THz and MW repetition frequency in a high-speed photodiode, nonlinear mixing of the photocurrents will directly yield an IF, taking advantage of MW harmonic generation in the photodiode. The origin of photodiode nonlinearity \cite{KJWilliams1990,KJWilliams1998,Tsuchiya1999,Fushimi2004} is still not fully understood. Several mechanisms contribute, such as voltage and current-dependence of responsivity and junction capacitance, space charge effect, Franz–Keldysh effect \cite{Andreas2008,huapu2010,Yangfu2011}. Photodiode nonlinearity is conventionally considered an adverse effect since it induces unwanted mixing products and results in reduced spurious-free dynamic range (SFDR) of the system. However, nonlinearity has been explored for optoelectronic mixing \cite{Hoshida1998,Tsuchiya1999,Fushimi2004,JWShi2009, Pan2010}. In those experiments, frequency up-conversion to V band has been demonstrated for applications in wireless communications. 

\begin{figure*}[t]
\centering
\includegraphics[width=1\linewidth,trim={0.2cm 0.6cm 0.1cm 1.15cm},clip]{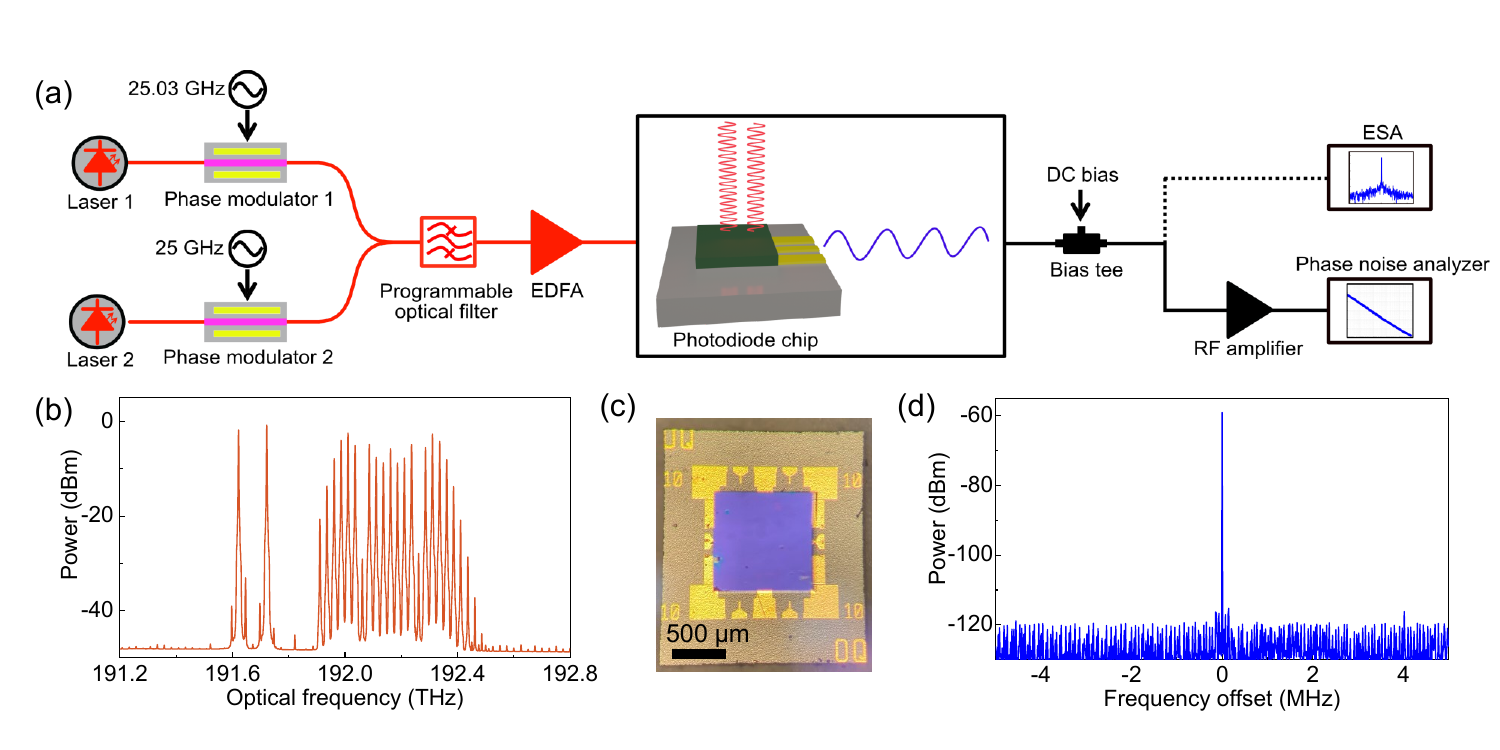}
\caption{Phase-coherent detection of $f_{\text{rep}}$ using a high-speed photodiode as the optoelectronic mixer. (a) Experimental setup, including two systems of lasers and phase modulators to generate EO combs, a programmable optical filter, an EDFA, the photodiode chip, a RF amplifier, an electrical spectrum analyzer, and a phase noise analyzer. (b) Optical spectrum at the output of the EDFA, showing a 100 GHz millimeter-wave-rate comb and a 25 GHz microwave-rate comb. (c) Microscope image of the photodiode chip after flip-chip bonding on aluminium nitride submount. (d) Electrical spectrum of the IF signal at 120 MHz.}
\label{Fig1}
\end{figure*}

Here, we draw on advances in high-speed MUTC photodiodes \cite{Qinglong2016,Jesse2018,Jesse2023} to demonstrate an ultra-broadband optoelectronic mixer operating at up to 500 GHz. Our photodiode mixer induces negligible frequency noise, allowing for Allan deviation (ADEV) frequency-noise measurements at $10^{-13}/\tau$ with the THz rate combs. We experimentally investigate the dependence of the conversion loss on photodiode's bias voltage and photocurrent, showing that an optimal bias voltage and large photocurrent is beneficial to minimizing the conversion loss. Featuring compact size, low additional noise and broad operating bandwidth, the optoelectronic mixer we report is a promising candidate for precise detection of frequency comb $f_{\text{rep}}$ at THz frequencies.

Our experimental setup is shown in Fig. 1(a), including the frequency-comb generation and the optoelectronic mixer systems. We use two electro-optic (EO) frequency combs with microwave repetition frequency ($ f_{\text{mw}}\approx 25 $ GHz) and a programmable optical filter to generate comb spectra with widely variable repetition frequency ($f_{\text{rep}}$) into the millimeter wave and THz ranges. Therefore, we can generate nearly arbitrary combs to explore the optoelectronic mixer. Operationally, we use two systems of continuous-wave (CW) lasers and phase modulators to generate two EO combs at 25.03 GHz and 25 GHz, respectively. We  tune the CW lasers across the C band, so that the center frequency of the EO combs can be very different. By programming an optical filter (Finisar Waveshaper 1000S), we create combs with arbitrary repetition frequency to characterize the optoelectronic mixer. We use an erbium doped fiber amplifier (EDFA) after the filter to increase the optical power for high-nonlinearity operation of the photodiode. The output of the photodiode is connected to a bias tee, which separates the DC bias voltage and IF signal. The IF signal is either fed to an electrical spectrum analyzer (ESA) for spectrum measurement or a phase noise analyzer for Allan deviation (ADEV) measurements. Figure 1(b) shows the case of creating a 100 GHz millimeter-wave-rate comb from the EO comb of laser 1 and a 25 GHz microwave-rate comb from the EO comb of laser 2.

In order to maximize the power of photo-generated electrical signal and IF signal, we design the photodiode with ultra-wide bandwidth and high power handling capability. We use a MUTC structure \cite{Qinglong2016,Jesse2018,Xie2024} with absorption layer thickness of 180 nm to increase the transit-time-limited bandwidth and saturation power. To minimize the junction capacitance for high-speed operation, we reduce the active area to 5 $\mu$m in diameter. Figure 1(c) presents a microscope image of our photodiode with 3 dB bandwidth of 128 GHz and a maximum photocurrent over 30 mA. The high-speed photodiode is flip-chip bonded on an aluminum-nitride submount. On the submount, there are coplanar waveguides (CPWs), which are compatible with commercial ground-signal-ground (GSG) microwave probes. The geometry of the CPWs is also optimized to further enhance the photodiode bandwidth through inductive peaking \cite{Gould2012, Novack2013}. 

\begin{figure}[htb]
\centering
\includegraphics[width=1\linewidth,trim={0.0cm 1.1cm 0cm 1.15cm},clip]{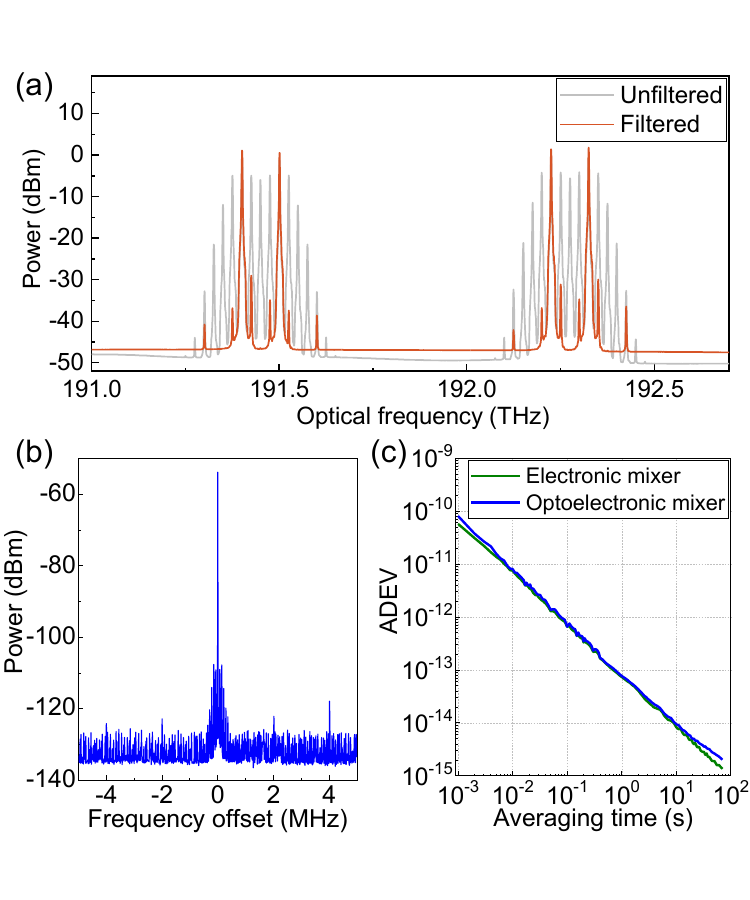}
\caption{A comparison of the noise performance between a conventional electronic mixer and  our optoelectronic mixer. (a) Optical spectra of the unfiltered (gray) and filtered (red) EO combs. (b) Electrical spectrum of the IF signal at 120 MHz. (c) A comparison of ADEV measured with a conventional electronic mixer and our optoelectronic mixer.}
\label{Fig2}
\end{figure}

Photodetection of both combs generates not only the harmonics of their repetition rates but also their inter-modulation products because of photodiode nonlinearity. In Fig. 1(d), we present the baseband electrical spectrum when the photodiode is illuminated with the optical spectrum of Fig. 1(b). Here, the central frequency is $f_\text{IF}=120$ MHz, corresponding to $f_\text{rep}$ of 100.12 GHz and the mixing product of $4 f_{\text{mw}}$, which is generated inside the photodetector by a combination of photocurrent harmonics and the intrinsic comb tones that lead to $4 f_{\text{mw}}$. One feature of our optoelectronic mixer is that no high-frequency components are required for THz detection. We can use low-frequency components for the IF signal and they also function as low-pass filters to filter out high-frequency signals.

We investigate the noise performance of the optoelectronic mixer in the millimeter-wave frequency range by operating it to measure the repetition frequency fluctuations of two, $\approx$100 GHz combs. Figure 2(a) shows how we generate the 100 GHz combs by programming the optical filter to transmit only two modes from each EO comb; the gray (red) trace shows the unfiltered (filtered) EO combs measured at the EDFA output. The resulting combs have spacings of 100 GHz and 100.12 GHz, leading to a down-mixed $\text{IF}$ signal of 120 MHz; see Fig. 2(b). The IF signal has a signal-to-noise ratio (SNR) over 60 dB, making high coherence phase and frequency noise analysis readily feasible. We measure the ADEV of the $\text{IF}$ signal, using a conventional phase noise analyzer with the same clock input as we use to generate the 25 GHz drive signal for the EO combs. Hence, the ADEV of the IF signal is traceable to that of the shared clock input, providing the baseline for characterizing the noise performance. We compare the optoelectronic mixer with a conventional electronic mixer by measuring the ADEV of the IF signal in both cases; see Fig. 2(c). In the case with the electronic mixer, 25 GHz and 25.03 GHz signals used for the EO combs are fed into an electronic mixer to generate the IF signal, which allows us to estimate the fundamental noise in down-mixing of 100 GHz combs. Figure 2(c) shows that the ADEV measured with the optoelectronic mixer agrees with that measured by a conventional electronic mixer, reaching $7.7 \times 10^{-14}$ at 1 second averaging time. These results indicate that the noise induced by our optoelectronic mixer is negligible compared with the conventional electronic mixer.

 \begin{figure}[htb]
\centering
\includegraphics[width=1\linewidth,trim={0cm 2.5cm 0cm 0cm},clip]{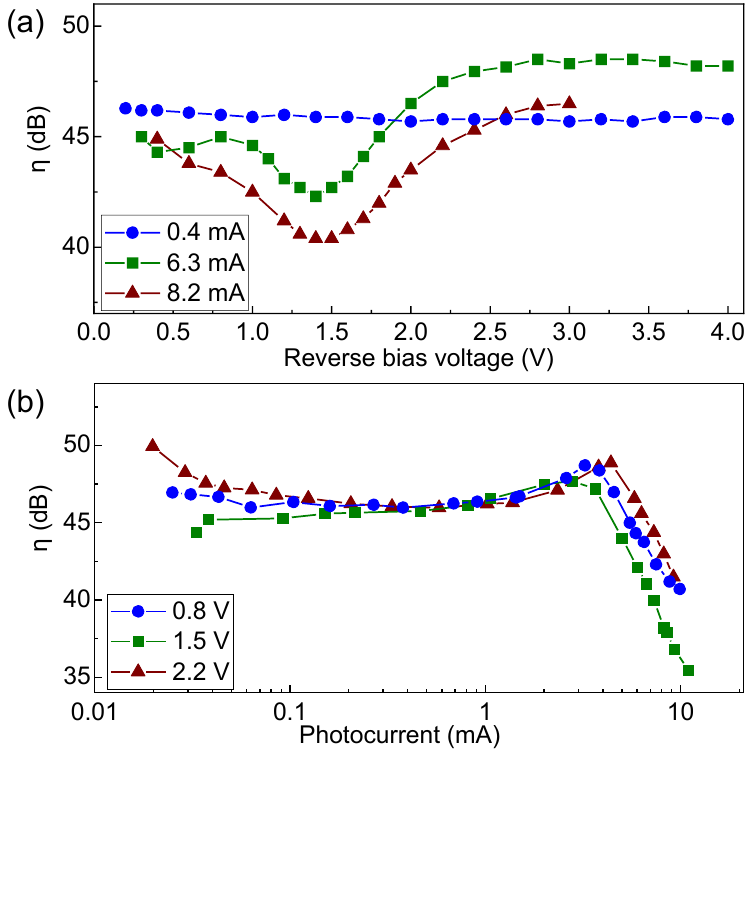}
\caption{Conversion loss of the optoelectronic mixer in the case of two, $\approx$100 GHz combs. (a) Measured conversion loss as a function of reverse bias voltage. (b) Measured conversion loss with varied photocurrent.}
\label{Fig3}
\end{figure}

Photodiode nonlinearity is essential for our optoelectronic mixer, and we optimize the photodiode nonlinearity by investigating the conversion loss, which is defined as $\eta=10\text{log}_{10} (P_{\text{mmW}}/ P_{\text{IF}})$, where $P_{\text{mmW}}$ and $P_{\text{IF}}$  are the power of photo-generated millimeter wave signal and IF signal, respectively \cite{JWShi2009}. For example, we measure $\eta$ in the case of two, $\approx$100 GHz combs. Since both photodiode's bias voltage and photocurrent have influence on its nonlinearity, we investigate $\eta$ by varying one parameter while keeping the other fixed. 

Figure 3(a) shows $\eta$ as a function of reverse bias voltage at different photocurrent.  At 0.4 mA photocurrent (blue circles), $\eta$ is independent of photodiode's bias voltage. At larger photocurrent (green squares and brown triangles ), $\eta$ first decreases and then increases with bias voltage, resulting in a minimum $\eta$ at reverse bias voltage of 1.4 V. This trend can be explained by the competition between higher responsivity and stronger internal electric field as the reverse bias voltage increases. It is believed that the photodiode nonlinearity is related to the accumulation of photo-generated carriers in the depletion region of the photodiode (space charge). The responsivity of the photodiode increases with the reverse bias voltage, leading to more space charge and higher nonlinearity. This effect is dominant at reverse bias voltages below 1.4 V. Above 1.4 V, another effect becomes dominant: the strong internal electric field in the photodiode sweeps out the space charge and reduces the photodiode nonlinearity. For the MUTC photodiode we use, the competition between these two effects results in a minimum $\eta$ at 1.4 V.

Figure 3(b) illustrates how $\eta$ varies with increasing photocurrent, showing a trend of initially increasing slightly and then decreasing sharply. This trend is explained by the space charge effect. The MUTC photodiode contains an un-depleted absorption region where the movement of photo-generated electrons is mainly based on diffusion at low photocurrent level. As the photocurrent increases, an electric field, which is called self-induced field, arises in that region \cite{Shimizu1998,Ishibashi2000}. It is beneficial to electrons' transport and leads to reduced space charge and lower nonlinearity. However, this electric field is relatively weak and the above-mentioned effect only dominates at low photocurrent level. At larger photocurrent, the accumulation of photo-generated carriers becomes the predominant effect and the photodiode nonlinearity is enhanced by the increased space charge. These experimental results suggest that we can operate the photodiode at as large photocurrent as possible to improve its nonlinearity. 

 \begin{figure}[htb]
\centering
\includegraphics[width=1\linewidth,trim={0.0cm 1.15cm 0cm 1.15cm},clip]{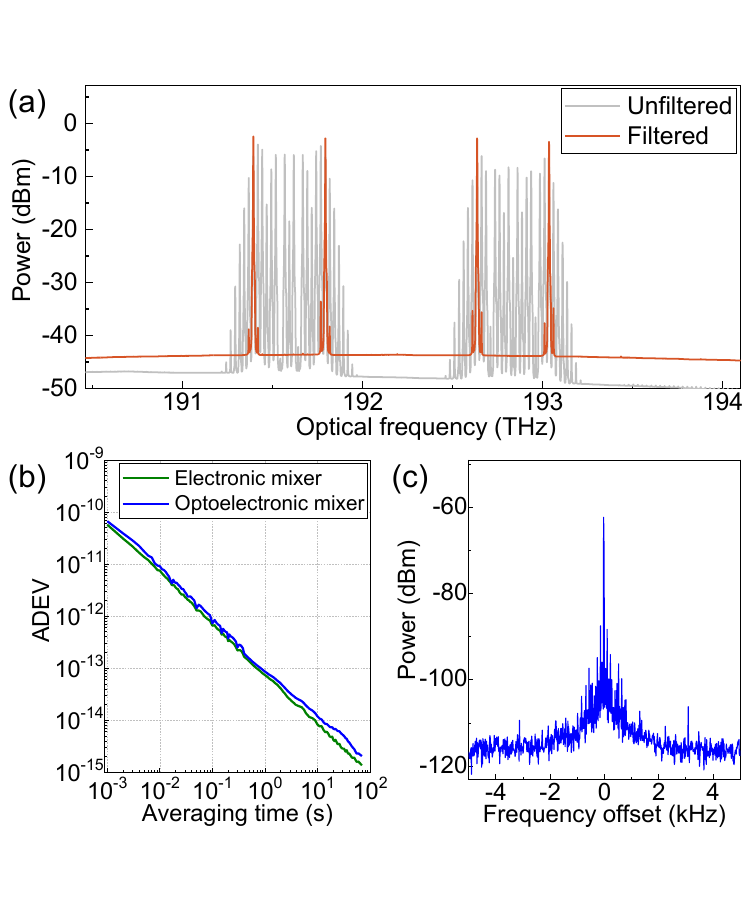}
\caption{Broadband frequency-noise performance of the optoelectronic mixer. (a) Optical spectrum (red) of two, $\approx$400 GHz combs. The gray trace represents the case of unfiltered EO combs. (b) A comparison of the ADEV measured by a conventional  electronic mixer and our optoelectronic mixer. (c) Electrical spectrum of the IF signal in the case with two, $\approx$500 GHz combs.}
\label{Fig4}
\end{figure}

We further investigate the broadband frequency-noise performance of our optoelectronic mixer by operating it with two, $\approx$400 GHz combs (Fig. 4 (a) and (b)) and two, $\approx$500 GHz combs (Fig. 4 (c)). Figure 4(a) shows the optical spectra in the case with (red) and without (gray) applying the multiple passband filter. By programming the optical filter, we achieve two combs with spacing of 400.12 GHz and 400 GHz, respectively. Photodetection of these two combs and down-mixing of the corresponding electrical signals yield $f_{\text{IF}}$ at 120 MHz. Figure 4(b) shows a comparison of ADEV between the case with the conventional electronic mixer and our optoelectronic mixer. In both cases, the ADEV is less than $8.6 \times 10^{-14}$ at 1 second averaging time. The IF signal generated by our optoelectronic mixer shows slightly higher frequency noise, which results from the noise induced by optical and electrical amplifiers in the system. We further operate the optoelectronic mixer with two, $\approx$500 GHz combs, see Fig. 4(c). The SNR  of the IF signal is low, partly due to the weak comb lines and low-SNR optical signals after the EDFA. We can improve the SNR of the IF signal by using cascaded modulators to generate 500 GHz combs with higher comb line power. Compared with the conventional electronic mixer, our optoelectronic mixer can operate over a much broader frequency bandwidth, supporting down-mixing of combs at millimeter wave and THz ranges.

In summary, we demonstrate a broadband optoelectronic mixer, based on a high-speed MUTC photodiode. Using a single photodiode, we demonstrate phase-coherent detection of $f_{\text{rep}}$ up to 500 GHz. The exceptional noise performance of the optoelectronic mixer is confirmed by the comparison with a conventional electronic mixer. We optimize the photodiode nonlinearity by investigating the dependence of  conversion loss on photodiode's bias voltage and photocurrent. To minimize the conversion loss, we can operate the photodiode at an optimal reverse bias voltage and maximize the photocurrent. Our work will benefit applications in optical frequency synthesis by providing a compact and broadband solution for detection of millimeter wave and THz repetition rates.

\section*{Acknowledgment}
We acknowledge Madison Woodson and Steven Estrella from Freedom Photonics for MUTC PD fabrication. We thank Haixin Liu, Nitesh Chauhan, and Matthew Hummon for technical review of the letter. This research has been funded by the AFOSR FA9550-20-1-0004 Project Number 19RT1019, DARPA A-PhI FA9453-19-C-0029, NSF Quantum Leap Challenge Institute Award OMA – 2016244, and NIST. This work is a contribution of the U.S. government and is not subject to copyright. Tradenames provide information only and not an endorsement.

\section*{Disclosures} The authors declare no conflicts of interest.
%\noindent \textbf{Disclosures.} The authors declare no conflicts of interest.

%\section{References}

\bibliography{sample}

\bibliographyfullrefs{sample}

\end{document}